# Intrinsic Strain-Driven Topological Evolution in SrRuO$_3$ via Flexural Strain Engineering

*Liguang Gong, Hongping Jiang, Bin Lao, Xuan Zheng, Xuejiao Chen, Zhicheng Zhong, Yan Sun, Xianfeng Hao,\* Milan Radovic, Run-Wei Li,\* Zhiming Wang\**

L. Gong, B. Lao, X. Zheng, R.-W. Li, Z. Wang
Ningbo Institute of Materials Technology and Engineering
Chinese Academy of Sciences
Ningbo 315201, China
E-mail: runweili@nimte.ac.cn; zhiming.wang@nimte.ac.cn

L. Gong, R.-W. Li, Z. Wang
Center of Materials Science and Optoelectronics Engineering
University of Chinese Academy of Sciences
Beijing 100049, China

H. Jiang, X. Hao
Key Laboratory of Applied Chemistry
College of Environmental and Chemical Engineering
Yanshan University
Qinhuangdao 066004, China
E-mail: xfhao@ysu.edu.cn

X. Chen
School of Photoelectric Engineering
Changzhou Institute of Technology
Changzhou, Jiangsu, 213002, China

Z. Zhong
Suzhou Institute for Advanced Research
School of Artificial Intelligence and Data Science
University of Science and Technology of China
Suzhou 215123, China

Y. Sun
Shenyang National Laboratory for Materials Science
Institute of Metal Research
Chinese Academy of Sciences

Shenyang 110016, China

M. Radovic

Swiss Light Source

Paul Scherrer Institut

CH-5232 Villigen PSI, Switzerland

Run-Wei Li

Eastern Institute of Technology

Ningbo 315200, China

Strain engineering offers a powerful route to tailor topological electronic structures in correlated oxides, yet conventional epitaxial strain approaches introduce extrinsic factors such as substrate-induced phase transitions and crystalline quality variations, which makes the unambiguous identification of the intrinsic strain effects challenging. Here, we develop a flexural strain platform based on van der Waals epitaxy and flexible micro-fabrication, enabling precise isolation and quantification of intrinsic strain effects on topological electronic structures in correlated oxides without extrinsic interference. Through strain-dependent transport measurements of the Weyl semimetal $SrRuO_3$, we observe a significant enhancement of anomalous Hall conductivity (AHC) by 21% under a tiny strain level of 0.2%, while longitudinal resistivity remains almost constant—a hallmark of intrinsic topological response. First-principles calculations reveal a distinct mechanism where strain-driven non-monotonic evolution of Weyl nodes across the Fermi level, exclusively governed by lattice constant modulation, drives the striking AHC behavior. Our work not only highlights the pivotal role of pure lattice strain in topological regulation but also establishes a universal platform for designing flexible topological oxide devices with tailored functionalities.

# 1. Introduction

Topological quantum materials have emerged as a revolutionary platform for next-generation technologies, ranging from energy-efficient electronics to fault-tolerant quantum computing, owing to their topological states protected by nontrivial electronic structures.[1-5] Among these materials, 4/5d transition metal oxides (TMOs), exhibit distinctive advantages in hosting exotic topological states.[6,7] The strong spin-orbit coupling (SOC) and electron correlations inherent to TMOs enable the realization of Dirac nodes,[8,9] magnetic Weyl nodes,[10,11] and other nontrivial topological electronic configurations,[12,13] where large intrinsic Berry curvature manifests as anomalous Hall conductivity (AHC) and spin-orbit torques (SOTs).[14-16] Moreover, the topological electronic structure of TMOs can be effectively modulated by external stimuli—including electric fields,[17] interface,[18] and strain[11,19]—due to the intricate interplay among lattice, spin, and orbital degrees of freedom in correlated oxide systems.[20,21] These unique characteristics establish TMOs as a versatile platform for both fundamental investigations of topological phenomena and the design of topological electronic devices.

Strain engineering is widely regarded as a highly effective and versatile approach for tailoring topological electronic structures by modulating lattice spacing and symmetry.[22,23] In TMOs, for instance, topological electronic structures and Berry curvature distributions in magnetic Weyl semimetal $SrRuO_3$ (SRO) and Dirac nodal line semimetal $SrIrO_3$ thin films have been significantly modulated via epitaxial strain control from various substrates, leading to sign reversal of AHC and enhancement of SOT.[19,24-27] However, some intertwined effects, such as structural phase transitions (e.g., orthorhombic-to-tetragonal transformations) and variations in crystalline quality (e.g., resistivity change),[26,28-30] are commonly introduced simultaneously into the TMOs based on the substrate-control strain engineering approach. These undesirable extrinsic factors make the identification of strain affect intrinsically to topological properties challenging. To decode this complexity, a pure lattice strain-modulation platform is urgently essential to disentangle strain effect from such extrinsic factors and establish unambiguous correlations between strain and topological electronic structure in TMOs.

In this work, we develop a flexural strain engineering approach to selectively induce strain while suppressing phase transitions and crystalline quality variations, and systematically investigate intrinsic strain effects on the topological state associated magnetoelectric properties in SRO thin films. Through van der Waals epitaxy on mica substrate and developed flexible micro-fabrication process, we obtain magnetoelectric transportation favoured Hall-bar devices of single crystalline SRO with stain tunability and high mechanical stability. We observe a 21% enhancement of AHC at a strain level of 0.2%, accompanied by an invariant longitudinal resistivity, unambiguously demonstrate a strain-induced topological structures evolution rather than extrinsic scattering mechanisms. First-principles calculations reveal that composition variations and energy level shifts of the $t_{2g}$ orbitals solely via lattice constant control can give rise to non-monotonic migration of Weyl nodes across the Fermi level, which is directly correlating with the experimentally observed AHC behavior. This approach isolates the exclusive role of pure lattice strain in modulating topology and AHE, providing direct evidence of its dominant influence over extrinsic factors in flexible oxide-based topological devices.

## 2. Results and Discussions

### 2.1. Epitaxial Growth and Fabrication of Flexible SrRuO3 Devices on Mica

A 20-nm (111)-oriented SrRuO3 film is epitaxially deposited on a SrTiO3 (STO)-buffered Mica substrate using pulsed laser deposition (PLD), as shown in Figure 1a. The STO buffer layer, with a lattice constant of 3.905 Å, closely matches the lattice constant of most perovskite oxides and serves as a dielectric material.[31] Moreover, the STO layer can be synthesized as a single-crystal film with a preferred (111) crystalline orientation on the mica substrate, as shown in Supporting Information Figure S1c, facilitating the growth of other perovskite oxides.[32,33] The epitaxial growth of the STO buffer layer on Mica is crucial for the successful fabrication of high-quality SRO films with atomically smooth surfaces. Benefitting from this, the (111)-oriented SRO was fabricated epitaxially with an atomically smooth surface exhibiting a roughness of 0.57 nm, as shown in Figure S1.

Next, a standard micro-fabricating process is employed to create a micron-sized Hall bar pattern on the Mica, as seen in Figure 1b. To ensure device flexibility, two additional steps - flexible support and mechanical exfoliation - are carried out. First, the device is adhered to a heat release tape and thinned down to 10 μm using PI tapes. This thinning process is essential for achieving the desired flexibility while maintaining the structural integrity of the epitaxial films. Subsequently, the device is detached from the heat release tape by baking it at 120 °C. This entire process yields a flexible and bendable SRO-based Hall-bar device, as demonstrated in Figure 1c, which can sustain a large bending curvature with a radius of 2 mm without generating any cracks or fractures.

### 2.2. Flexural Strain-Induced Modulation of Anomalous Hall Effect

The successful fabrication of flexible microdevices on mica substrates unlocks the potential for applying tunable and reversible strain without inducing unwanted structural phase changes, thus offering a powerful platform to probe the intrinsic influence of strain on the anomalous Hall effect (AHE) in SRO. Figure 2a presents a schematic of AHE measurement setup, featuring an out-of-plane magnetic field ($B$), applied current ($I$) and acquired voltage signals on our flexible Hall-bar devices. During bending measurements, the applied flexural strain ($\varepsilon$) on the Hall-bar devices is regulated by changing the curvature radius ($R$) of the Teflon mold. The corresponding strain is calculated using the formula $\varepsilon = \pm t/2R$, where $t$ is the total thickness of mica substrate and film, and + (-) represents tensile (compressive) uniaxial in-plane strain, respectively. Moreover, the deviation in the magnetic field due to the curved profile of the microdevices is negligible and does not impact the AHE measurements, as demonstrated in Figure S2 and Supporting Information.

To establish a baseline of the strain modulation, we first measure the temperature dependence of AHE in the flat state. Figure 2b shows the anomalous Hall resistivity $\rho_{xy}^{AHE}$ as a function of magnetic filed at different temperature. At lower temperature regime, a distinct square-like hysteresis loop emerges, with the magnitude initially increasing before gradually decreasing as the temperature rises. However, once the temperature surpasses 140 K, the hysteresis features are notably suppressed and undergo a change in sign. With further temperature elevation, the AHE diminishes once more until it ultimately vanishes beyond the Curie temperature ($T_c$ = 159 K, Figure S1d). This non-monotonic temperature dependence of the AHE in SRO aligns with previous studies and has been primarily attributed to intrinsic mechanisms and the topological electronic structures.[14]

To elucidate the influence of strain effect on the AHE, we systematically measured the AHE loops as a function of strains ranging from -0.2% to +0.25%, which control by the curvature radius ($R$) of the mold. Figure 2c illustrates the AHE loops at 60 K under various strain states, revealing that both compressive and tensile strains enhance the amplitude of AHE with respect to the strain-free state. Remarkably, this enhancement exhibits an asymmetric behavior, while a mere 0.2% compressive strain boosts $\rho_{xy}^{AHE}$ by an impressive 21% (from 0.451 to 0.544 µΩ·cm), far surpassing the 9% maximum enhancement observed under tensile strain. This pronounced asymmetry likely stems from the anisotropic response of SRO's topological band structure to lattice distortion, suggesting a directional dependence of strain-induced Berry curvature modulation that warrants further exploration. Strain-dependent AHE measurement were also conducted at different temperatures, as shown in Figure S3, with the corresponding results of anomalous Hall conductivity (AHC) $\sigma_{xy}^{AHE} = -\rho_{xy}^{AHE} / \rho_{xx}^2$ (where $\rho_{xx}$ is the longitudinal resistivity) plotted as a function of temperature and strain in Figures 2d and 2e, respectively. The results reveal that the non-monotonic temperature dependence of AHC is preserved across all strain states, with the strain-induced enhancement exhibiting a similar asymmetric trend below 120 K, most pronounced at 60 K where $\sigma_{xy}^{AHE}$ reaches its maximum absolute value. This consistency underscores that the observed AHC is protected by the topological nature of the SRO band structure raising from magnetism, with strain and temperature merely modulating its magnitude.

Apart from the significant change of AHC, intriguingly, the simultaneously measured longitudinal resistivity $\rho_{xx}$ under the strain modulation remains almost unchanged, as depicted in Figure 2f. This fact reflects that the applied strain is insufficient to induce an obviously change of lattice scattering. Namely, the resistivity dependent factors, such as concentration and distribution of impurities, structural phase transitions, and lattice defects, are effectively suppressed during strain application. As shown in Figure 2g, while the maximum change in $\sigma_{xy}^{AHE}$ exceeds 21%, the change in $\rho_{xx}$ under the same strain is less than 1% at 60 K. This stark contrast between substantial AHC modulation (up to 21%) and the negligible variation in $\rho_{xx}$ (<1%) provides compelling evidence that the observed strain modulated AHC behavior is purely intrinsic, independent of extrinsic mechanisms such as side-jump or skew scattering.[15] This finding not only confirms the topological origin of AHE in SRO but also highlights the unparalleled efficacy of pure lattice strain in modulating intrinsic Berry curvature, establishing a cornerstone for designing topology-driven devices with tailored functionalities.

## 2.3. Topological Electronic Structures, Berry Curvature and AHC

To further investigate the role of topological electronic structure in the modulation of AHC, we conducted first-principles calculations on the orthorhombic phase of SRO (Methods and Supporting Information). By considering the spin-orbit coupling (SOC) and aligning the magnetization along the orthorhombic *b*-axis, we identified all band crossing points near the Fermi level within the first Brillouin zone, as illustrated in Figure 3a. Among these, certain isolated points with opposite chiralities (blue and red representing positive and negative chiralities, respectively) are referred to as Weyl nodes. These Weyl nodes are predominantly distributed in the $k_a = 0$ and $k_c = 0$ planes (see Supporting Information Table S1 for more details), consistent with previous reports.[10,17] The four Weyl nodes closest to the Fermi level, located at (0, ±0.457, ±0.389) with an energy of -13 meV, forming two pairs with opposite signs of momentum. The pair of Weyl nodes at (0, 0.457, 0.389) and (0, -0.457, -0.389) (highlighted with colored circles in Figure 3a) have their three-dimensional energy dispersion projected onto the $k_a - k_{\parallel, N}$ plane, where $k_{\parallel, N}$ runs parallel to the direction connecting these two Weyl nodes, as shown in Figure 3b. The band structure near the Weyl nodes is tilted, indicative of type-II Weyl nodes,[34] as further evidenced by touching points between the electron and hole pockets in the Fermi surface (see Figure S5). In addition to the Weyl nodes, a loop formed by continuous crossing points appears on the $k_b = 0$ plane, referred to as nodal line,[35,36] depicted by the black line in Figure 3a. The three-dimensional energy dispersion of the nodal line, projected onto the $k_a - k_c$ plane (Figure 3c), reveals that the two bands near the Fermi level touch to form a nodal line loop, which intersects the Fermi surface at eight distinct points.

As topological electronic structures, Weyl nodes and nodal line are expected to contribute to the Berry curvature and observable AHC when they are close to the Fermi level.[37-39] To quantify their contributions, we computed the band structure and Berry curvature of the Weyl nodes and nodal line by tracing paths that connect either the pair of Weyl nodes or two intersections (black circles in Figure 3c) of the nodal line, as depicted in Figure 3d and 3e. Weyl nodes of opposite chiralities (circled in color) exhibit large Berry curvature $\Omega_z$, with a sharp peak of -2200 Å$^2$ near them. In contrast, the Berry curvature near the nodal lines is lower and more broadly distributed, with a positive maximum value of 60 Å$^2$ along the path. This highlights the dominant contribution of Weyl nodes to the overall Berry curvature. The reduced contribution from the nodal line could be attributed to other crossing points surrounding the eight intersections, which are distributed on both sides of the Fermi surface, leading to partial cancellation of their curvature contributions. To evaluate of the impact of these topological electronic structure on the AHE, we calculated the AHC in SRO as a function of energy (Figure 3f). Near the Fermi level, we obtained a substantial AHC of -430 $\Omega^{-1}\cdot cm^{-1}$, demonstrating the significant contribute of the topological electronic structures, particularly the Weyl nodes, to the intrinsic AHE in the orthorhombic phase of SRO.

## 2.4. Strain-Induced Modulation of Weyl Nodes and AHC

To elucidate the mechanism behind the experimentally observed non-monotonic strain modulations of the AHC in SRO, we performed a comprehensive analysis of the band structure and its evolution under strain. In transition metal oxide SRO, the electronic states near the Fermi level are predominantly derived from the nearly degenerate Ru-$t_{2g}$ orbitals (Figure S4b and S6a). Strain alters the distances and angles between the Ru and oxygen ions in SRO, thereby modifying the crystal field around the Ru ions and resulting in the energy level splitting of Ru-$t_{2g}$ bands.[19,24] The flexural mica substrate used in our experiments induces a remarkably clean and subtle strain, free from the complexities of phase transition commonly associated with epitaxial strain approaches, ensuring a pure lattice deformation effect.[19,40] To simulated the effect of flexural strain in our calculations, we adjusted the lattice constant along one crystal axis. As illustrated in Figure 4a, increasing (or decreasing) the lattice parameter $c$ stretches (or compresses) the Ru-O octahedra, leading to shifts in the energy levels of the three $t_{2g}$ orbitals. These energy shifts, in turn, influences the energy positions and the orbital composition of the Weyl nodes, which play a crucial role in modulating the AHC.

To quantify the energy level splitting of Ru-$t_{2g}$ bands, we analyzed the energy distribution of the $t_{2g}$ density of states (DOS) (see details in Figure S6). We used the integrated DOS of the three orbitals up to the Fermi level at zero strain as a reference. For the strained states, we integrated the DOS up to this reference value, with the corresponding energy difference representing the orbital shift. As illustrated in Figure 4b, the $d_{xy}$ orbital shifts to higher energies as strain increases from negative to positive, while the $d_{yz}$ orbital shifts to lower energies. However, the $d_{xz}$ orbital exhibits a more complex behavior, initially shifting to higher energies before decreasing after a certain positive strain threshold. To validate these findings, we employed an alternative method (Figure S7), tracking the strain-induced shifts in the DOS peaks of the three orbitals near the Fermi level, which corroborated the trends observed in Figure 4b. Furthermore, by projecting the orbitals onto the band structure, we analyzed the orbital composition at the Weyl nodes. As depicted in Figure 4c, the color and size of the spheres represent the orbital composition and intensity, respectively. At -1% and 0% strain, the two bands forming the Weyl nodes are primarily composed of the $d_{xy}$ and $d_{xz}$ orbitals. However, under 1% strain, the $d_{yz}$ orbital gradually replaces the $d_{xy}$ orbital, along with the $d_{xz}$ orbital, becoming the dominant orbital composition at the Weyl nodes. Strikingly, this strain-induced transformation in Weyl node orbital composition emerges as a pivotal factor in unraveling the non-monotonic behavior of the AHC, as it fundamentally alters the electronic topology and Berry curvature distribution near the Fermi level.

To gain deeper insights into the strain-induced modulation of the AHC, we investigated the evolution of the Weyl nodes in momentum and energy space under different strains conditions. Remarkably, within the ±1% strain range, our calculations consistently identified the Weyl nodes near their initial positions, highlighting the topological protection and high stability of this pair of Weyl nodes in SRO. The Weyl nodes exhibited no gap opening and only slight shifts in momentum and energy as shown in Figure S8, where strain induces small shifts in $k_b$ and $k_c$ directions, but the Weyl nodes remain confined to the $k_b = 0$ plane. The position of the Weyl nodes relative to the Fermi level is closely related to the Berry curvature and the magnitude of the AHC. Figure 4d reveals the intricate energy shift of the Weyl nodes under varying strain. As strain increases, the Weyl node gradually shifts to higher energy, crossing the Fermi level near 0% strain, followed by a sudden jump at approximately 0.15% strain, reaching their maximum energy. With further strain increases, the Weyl node energy decreases and crosses the Fermi level again near 1% strain. This non-monotonic energy shift, a hallmark of strain-driven topological modulation, can be attributed to the intricate interplay of strain-induced energy and composition changes in the $t_{2g}$ orbitals. Under compressive strain, the Weyl node forms at the crossing of two bands dominated by the $d_{xy}$ and $d_{xz}$ orbitals, whose energies decrease, leading to a downward shift in Weyl node energy. Conversely, under tensile strain, the $d_{yz}$ orbital replaces $d_{xy}$, and the Weyl nodes are primarily composed of the $d_{xz}$ and $d_{yz}$ orbitals, which also decrease in energy, resulting in another downward shift. Together, these findings unveil a novel mechanism of strain-driven Weyl node evolution, offering a new dimension for tailoring topological properties in correlated oxides.

Given that Weyl nodes act as primary sources of Berry curvature and AHC, their non-monotonic migration across the Fermi level inevitably leads to complex AHC behavior. To quantify the influence of Weyl node migration, we calculated the AHC of SRO under different strain conditions, as shown in Figure 4e. As strain increases from -1% to 1%, the magnitude of AHC exhibits a non-monotonic change and reaches a minimum value around strain-free state, which is consistent with our experimental results presented in Figure 2g. These results indicate that, from a microscopic perspective, the modulation of AHC via strain effect is essentially influencing the movement of the topological band structures near Fermi surface. Moreover, the complexity of this non-monotonic AHC behavior is further compounded by the intricate interplay with other topological structures, such as nodal lines and additional Weyl nodes (Table S1, as discussed in Figure S9 and S10). While these structures exhibit irregular energy shifts under strain, their contributions remain secondary to the dominant Weyl nodes closest to the Fermi level, reinforcing the central role of the latter in driving the observed AHC response. It is worth emphasizing here that, firstly, our calculate results are only take into account the strain effect on lattice constant without any extrinsic factor such as octahedral rotation/tilting and defect. Secondly, as we described above, our developed experimental approach involves strain induced lattice compression or tension, neither includes substrate-induced phase transitions nor resistivity change. Therefore, the excellent agreement between our experimental and calculated $\sigma_{xy}^{AHE}$ unequivocally demonstrate that AHC modulation in our flexural strain platform is governed by intrinsic strain effects on topological electronic structures. This stands in stark contrast to conventional epitaxial strain methods, where extrinsic factors like structural transition and lattice scattering obscure the pure strain response.

## 3. Conclusion

In summary, we develop a strain modulation platform for topological oxides without introducing undesirable extrinsic factors, and subsequently demonstrate the strain control is effectvie to topological electronic structures and anomalous Hall effect of Weyl semimetal SRO. As the longitudinal resistivity remains almost unchanged, a significant non-monotonic behavior of the AHC with a 21% enhancement under a strain level of 0.2%, strongly suggesting an intrinsic origin related to the topological electronic structure. First-principles calculations indicate that $t_{2g}$ orbital composition variation and energy level shifts are dependent sensitively on intrinsic strain effect without structural phase transition or defect scattering, resulting in Weyl nodes evolution across the Fermi level and non-monotonic change of AHC, which is consistent with the experiment results. Our findings demonstrate that intrinsic lattice strain—distinct from phase transitions or resistance artifacts—directly drives topology and AHE evolution in correlated oxides, highlighting its irreplaceable role in designing flexible topological devices.

## 4. Experimental Section

*Film Growth and Device Fabrication*: The SrRuO$_3$ (20 nm) / SrTiO$_3$ (15 nm) bilayer samples were deposited on mica (001) substrates using pulsed laser deposition (PLD) with a KrF excimer laser ($\lambda = 248$ nm). The STO buffer layer was fist grown at 850 °C under an oxygen pressure of 0.15 mbar to improve the quality of the epitaxial films. Subsequently, the SRO film layer was deposited on the top of the buffer layer at 700 °C and 0.15 mbar oxygen pressure. Following SRO growth, the film was cooled to room temperature with a rate of 30 °C/min in a 0.15 mbar oxygen atmosphere.

For bending AHE measurement, the samples were patterned into Hall-bar devices using standard photolithography and Ar ion etching techniques. Here, the current path had dimensions of 75 μm in width (*W*) and 150 μm in length (*L*) between two adjacent voltage paths. Though the flexible process mentioned above, the mica substrates were tinned to 10 μm and attached to Teflon molds with different curvature radius.

*Film Characterization and AHE Measurement*: The surface appearance and thickness of the SRO films were assessed using in-situ reflection high-energy electron diffraction (RHEED) during growth and atomic force microscopy (AFM) scans post-growth. The magnetization and crystal structure were confirmed by superconducting quantum interference device (SQUID, Quantum Design) and X-ray diffraction (XRD). The AHE and resistance measurements were conducted using Keithley 6221 current source and SR830 lock-in amplifier within an Oxford cooling system equipped with a superconducting magnet.

*DFT Calculations*: As implemented in the Vienna Ab Initio Simulation Package (VASP), we performed first-principles DFT calculations using the generalized gradient approximation (GGA) with Perdew-Burke-Ernzerhof theory (PBE-sol) and the projection augmented wave method.[41-44] we constructed the model of orthorhombic SrRuO$_3$ structure, with crystallographic space group *Pnma* and lattice constants *a* = 5.514 Å, *b* = 5.551 Å, *c* = 7.802 Å,[17] comprising a total of 20 atoms: 4 Sr, 4 Ru, and 12 O atoms, resulting in a unit cell volume of 238.8 Å$^3$. Based on previous experimental observations, the magnetization direction was set along the *b*-axis.[17] Monkhorst-Pack *k*-point grids of 7×7×5 were used for Brillouin zone integration, with an energy cutoff of 500 eV. Considering the large atomic number and significant spin-orbit coupling of Ru atoms in SRO, spin-orbit coupling was included in the electronic property calculations. Additionally, the electronic correlation effects of Ru-$d$ orbitals were taken into account using the GGA+U method, with a Hubbard *U* value of 2 eV.

*Topological Properties and AHC Calculation*: To calculate the topological properties and AHC of SRO, we used the Wannier90 program to construct tight-binding Hamiltonians.[45,46] These Hamiltonians included Ru-4$d$ orbitals ($d_{xy}$, $d_{yz}$, $d_{xz}$, $d_{x^2-y^2}$ and $d_{z^2}$) and O-2$p$ orbitals ($p_x$, $p_y$ and $p_z$), comprising a total of 112 bands of maximally localized Wannier functions (MLWFs). The band structures obtained from the MLWFs tight-binding model were in good agreement with those from first-principles calculations (see Figure S4). Gaussian projection matrices and information on MLWFs were then written into Wannier90, and combined with input files for Wanniertools, enabling the calculation of topological properties.[47] The *k*-point grid size for integration calculations was typically set to 41×41×41 for identifying topological structures and chirality, and 301×301×301 for calculating band structures and Fermi surfaces.

*Model Construction Under Strain*: To investigate the effects of strain on the topological band structure, Berry curvature, and AHC, we constructed models of a single orthorhombic SRO unit cell under different strain conditions. The lattice constants *a* and *b* were kept fixed, while the strain was varied from -1% to +1% by adjusting the lattice constant c. No structural optimization was performed during the DFT calculations. The applied strain ($\varepsilon$) was calculated using the following Equation:

$$\varepsilon = \frac{c - c_0}{c_0} \times 100\% \tag{1}$$

where *c* is the lattice constant after stretching or compression, and $c_0$ represents the initial lattice constant without strain.

**Supporting Information**

Supporting Information is available from the Wiley Online Library or from the author.

**Acknowledgements**

This work was supported by the National Key Research and Development Program of China (Nos. 2024YFA1410200, 2019YFA0307800), the National Natural Science Foundation of China (Nos. 12174406, 11874367, 51931011, 52127803, 12274369), the Chinese Academy of Sciences Project for Young Scientists in Basic Research (No.YSBR-109), the Key Research Program of Frontier Sciences, Chinese Academy of Sciences (No. ZDBS-LY-SLH008), K.C.Wong Education Foundation (GJTD-2020-11), the Ningbo Key Scientific and Technological Project (Grant No. 2022Z094), the Central Guidance Local Science and Technology Development Fund Project No.246Z1104G.

**Conflict of Interest**

The authors declare that they have no conflict of interest.

**Data Availability Statement**

The data that support the findings of this study are available from the corresponding author upon reasonable request.

References

[1]  A. A. Burkov, *Nat. Mater.* **2016**, *15*, 1145.

[2]  P. Narang, C. A. C. Garcia, C. Felser, *Nat. Mater.* **2020**, *20*, 293.

[3]  B. A. Bernevig, C. Felser, H. Beidenkopf, *Nature* **2022**, *603*, 41.

[4]  O. Breunig, Y. Ando, *Nat. Rev. Phys.* **2022**, *4*, 184.

[5]  K.-H. Jin, W. Jiang, G. Sethi, F. Liu, *Nanoscale* **2023**, *15*, 12787.

[6]  M. Uchida, M. Kawasaki, *J. Phys. D: Appl. Phys.* **2018**, *51*, 143001.

[7]  S. G. Jeong, J. Y. Oh, L. Hao, J. Liu, W. S. Choi, *Adv. Funct. Mater.* **2023**, *33*, 2301770.

[8]  Z. T. Liu, M. Y. Li, Q. F. Li, J. S. Liu, W. Li, H. F. Yang, Q. Yao, C. C. Fan, X. G. Wan, Z. Wang, D. W. Shen, *Sci. Rep.* **2016**, *6*, 30309.

[9]  J. Fujioka, R. Yamada, M. Kawamura, S. Sakai, M. Hirayama, R. Arita, T. Okawa, D. Hashizume, M. Hoshino, Y. Tokura, *Nat. Commun.* **2019**, *10*, 362.

[10] K. Takiguchi, Y. K. Wakabayashi, H. Irie, Y. Krockenberger, T. Otsuka, H. Sawada, S. A. Nikolaev, H. Das, M. Tanaka, Y. Taniyasu, H. Yamamoto, *Nat. Commun.* **2020**, *11*, 4969.

[11] Y. Li, T. Oh, J. Son, J. Song, M. K. Kim, D. Song, S. Kim, S. H. Chang, C. Kim, B.-J. Yang, T. W. Noh, *Adv. Mater.* **2021**, *33*, 2008528.

[12] A. Shitade, H. Katsura, J. Kuneš, X.-L. Qi, S.-C. Zhang, N. Nagaosa, *Phys. Rev. Lett.* **2009**, *102*, 256403.

[13] B. Yan, M. Jansen, C. Felser, *Nat. Phys.* **2013**, *9*, 709.

[14] Z. Fang, N. Nagaosa, K. S. Takahashi, A. Asamitsu, R. Mathieu, T. Ogasawara, H. Yamada, M. Kawasaki, Y. Tokura, K. Terakura, *Science* **2003**, *302*, 92.

[15] N. Nagaosa, J. Sinova, S. Onoda, A. H. MacDonald, N. P. Ong, *Rev. Mod. Phys.* **2010**, *82*, 1539.

[16] P. Jadaun, L. F. Register, S. K. Banerjee, *Proc. Natl. Acad. Sci. U.S.A.* **2020**, *117*, 11878.

[17] W. Lin, L. Liu, Q. Liu, L. Li, X. Shu, C. Li, Q. Xie, P. Jiang, X. Zheng, R. Guo, Z. Lim, S. Zeng, G. Zhou, H. Wang, J. Zhou, P. Yang, Ariando, S. J. Pennycook, X. Xu, Z. Zhong, Z. Wang, J. Chen, *Adv. Mater.* **2021**, *33*, 2101316.

[18] R. Mori, P. B. Marshall, K. Ahadi, J. D. Denlinger, S. Stemmer, A. Lanzara, *Nat. Commun.* **2019**, *10*, 5534.

[19] D. Tian, Z. Liu, S. Shen, Z. Li, Y. Zhou, H. Liu, H. Chen, P. Yu, *Proc. Natl. Acad. Sci. U.S.A.* **2021**, *118*, e2101946118.

[20] E. Dagotto, *Science* **2005**, *309*, 257.

[21] C. Ahn, A. Cavalleri, A. Georges, S. Ismail-Beigi, A. J. Millis, J.-M. Triscone, *Nat. Mater.* **2021**, *20*, 1462.

[22] D. Du, J. Hu, J. K. Kawasaki, *Appl. Phys. Lett.* **2023**, *122*, 170501.

[23] J. M. Kim, M. F. Haque, E. Y. Hsieh, S. M. Nahid, I. Zarin, K.-Y. Jeong, J.-P. So, H.-G. Park, S. Nam, *Adv. Mater.* **2023**, *35*, 2107362.

[24] K. Samanta, M. Ležaić, S. Blügel, Y. Mokrousov, *J. Appl. Phys.* **2021**, *129*, 093904.

[25] J. Wei, H. Zhong, J. Liu, X. Wang, F. Meng, H. Xu, Y. Liu, X. Luo, Q. Zhang, Y. Guang, J. Feng, J. Zhang, L. Yang, C. Ge, L. Gu, K. Jin, G. Yu, X. Han, *Adv. Funct. Mater.* **2021**, *31*, 2100380.

[26] J. Zhou, X. Shu, W. Lin, D. F. Shao, S. Chen, L. Liu, P. Yang, E. Y. Tsymbal, J. Chen, *Adv. Mater.* **2021**, *33*, 2007114.

[27] Q. Zhang, S. Shi, Z. Zheng, H. Zhou, D.-F. Shao, T. Zhao, H. Su, L. Liu, X. Shu, L. Jia, Y. Gu, R. Xiao, G. Wang, C. Zhao, H. Li, J. Chen, *ACS Appl. Mater. Interfaces* **2024**, *16*, 1129.

[28] A. Vailionis, W. Siemons, G. Koster, *Appl. Phys. Lett.* **2008**, *93*, 051909.

[29] D. Kan, Y. Shimakawa, *Cryst. Growth Des.* **2011**, *11*, 5483.

[30] W. Lu, W. Song, P. Yang, J. Ding, G. M. Chow, J. Chen, *Sci. Rep.* **2015**, *5*, 10245.

[31] D. G. Schlom, L.-Q. Chen, C. J. Fennie, V. Gopalan, D. A. Muller, X. Pan, R. Ramesh, R. Uecker, *MRS Bull.* **2014**, *39*, 118.

[32] Y.-H. Chu, *npj Quantum Mater.* **2017**, *2*, 67.

[33] L. Lu, Y. Dai, H. Du, M. Liu, J. Wu, Y. Zhang, Z. Liang, S. Raza, D. Wang, C.-L. Jia, *Adv. Mater. Interfaces* **2020**, *7*, 1901265.

[34] A. A. Soluyanov, D. Gresch, Z. Wang, Q. Wu, M. Troyer, X. Dai, B. A. Bernevig, *Nature* **2015**, *527*, 495.

[35] A. A. Burkov, M. D. Hook, L. Balents, *Phys. Rev. B* **2011**, *84*, 235126.

[36] B. Q. Lv, T. Qian, H. Ding, *Rev. Mod. Phys.* **2021**, *93*, 025002.

[37] G. Chang, S.-Y. Xu, H. Zheng, B. Singh, C.-H. Hsu, G. Bian, N. Alidoust, I. Belopolski, D. S. Sanchez, S. Zhang, H. Lin, M. Z. Hasan, *Sci. Rep.* **2016**, *6*, 38839.

[38] P. Li, J. Koo, W. Ning, J. Li, L. Miao, L. Min, Y. Zhu, Y. Wang, N. Alem, C.-X. Liu, Z. Mao, B. Yan, *Nat. Commun.* **2020**, *11*, 3476.

[39] B. Sohn, E. Lee, S. Y. Park, W. Kyung, J. Hwang, J. D. Denlinger, M. Kim, D. Kim, B. Kim, H. Ryu, S. Huh, J. S. Oh, J. K. Jung, D. Oh, Y. Kim, M. Han, T. W. Noh, B.-J. Yang, C. Kim, *Nat. Mater.* **2021**, *20*, 1643.

[40] Y. Li, P. Zhou, Y. Qi, T. Zhang, *J. Am. Ceram. Soc.* **2021**, *105*, 2038.

[41] P. E. Blöchl, *Phys. Rev. B* **1994**, *50*, 17953.

[42] G. Kresse, J. Furthmüller, *Comput. Mater. Sci.* **1996**, *6*, 15.

[43] G. Kresse, J. Furthmüller, *Phys. Rev. B* **1996**, *54*, 11169.

[44] J. P. Perdew, K. Burke, M. Ernzerhof, *Phys. Rev. Lett.* **1996**, *77*, 3865.

[45] A. A. Mostofi, J. R. Yates, Y.-S. Lee, I. Souza, D. Vanderbilt, N. Marzari, *Comput. Phys. Commun.* **2008**, *178*, 685.

[46] N. Marzari, A. A. Mostofi, J. R. Yates, I. Souza, D. Vanderbilt, *Rev. Mod. Phys.* **2012**, *84*, 1419.

[47] Q. Wu, S. Zhang, H.-F. Song, M. Troyer, A. A. Soluyanov, *Comput. Phys. Commun.* **2018**, *224*, 405.

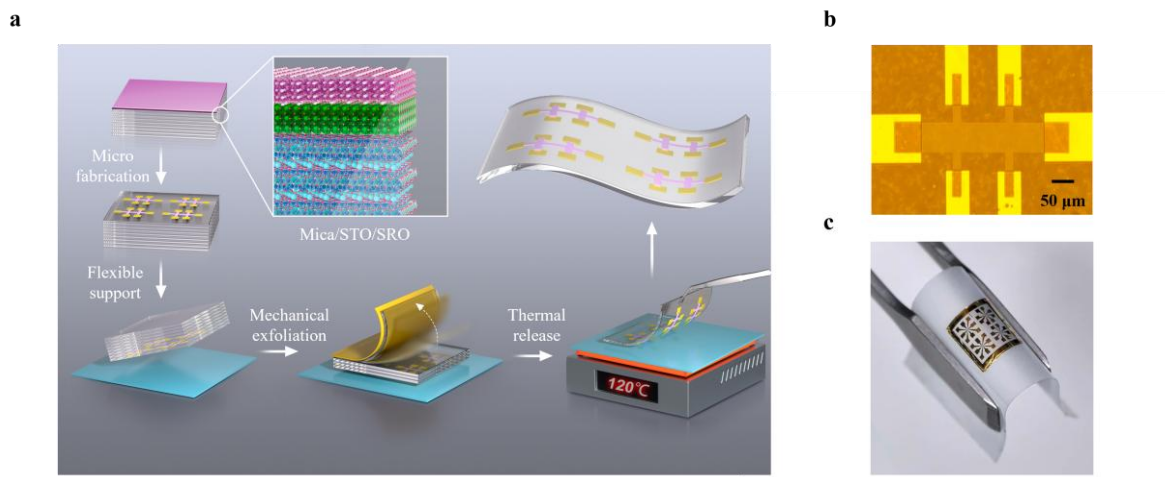

**Figure 1.** Fabrication of flexible oxide films and microdevices on mica substrates. a) Schematic diagrams illustrating the steps involved in the preparation of flexible device, from the epitaxial growth of the STO buffer layer and SRO film to the microfabrication of the Hall-bar pattern and the thinning process. b, c) Photographs of the flexible micro Hall-bar device, showcasing its bendability and mechanical stability.

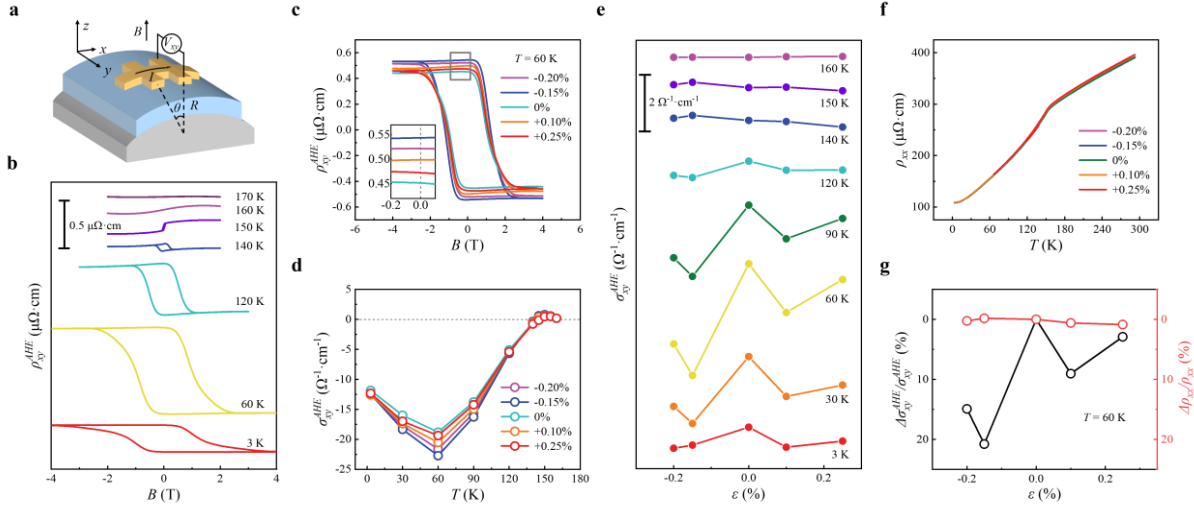

**Figure 2.** Modulation of AHE by flexural strain in SRO devices. a) Schematic illustration of AHE measurement setup on a flexible SRO-based Hall-bar device, with the magnetic field applied along the z-axis, and current injected along the strain direction. The applied strain is controlled by changing the curvature radius ($R$) of the Teflon mold. b) Anomalous Hall resistivity $\rho_{xy}^{AHE}$ as a function of magnetic field at various temperatures for the flat SRO device. c) Anomalous Hall resistivity loop at 60 K under varying flexural strains ranging from -0.2% to +0.25%. The inset provides a magnified view of the gray box near zero magnetic field. d) Temperature dependence of AHC $\sigma_{xy}^{AHE}$ under different flexural strains. e) The anomalous Hall conductivity as a function of flexural strain at different temperatures. f) The resistivity-temperature curves under different strains. The resistivity is almost independent of flexural strain, indicating that extrinsic mechanisms associated with scattering are not the primary factors in the observation of AHE. g) Comparison of the percentage change in AHC and device resistance as a function of flexural strain at 60 K, with the ordinates flipped for consistency due to the negative AHC values.

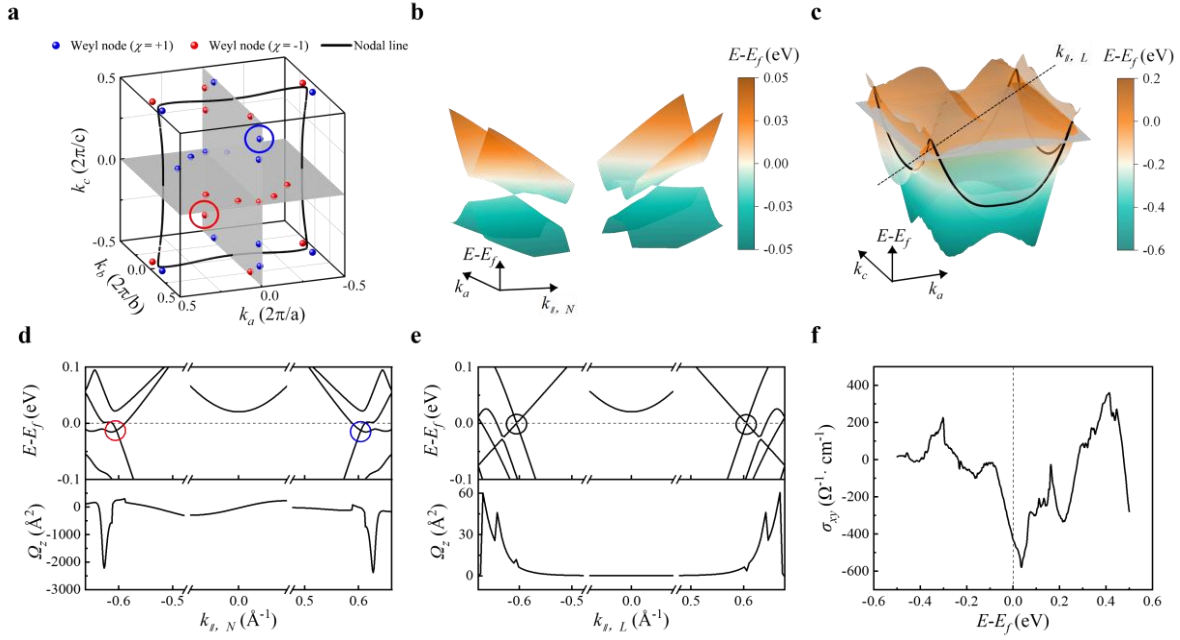

**Figure 3.** Topological electronic structures, Berry curvature and AHC in orthorhombic SRO. a) Distribution of Weyl nodes and a nodal line in the Brillouin zone of SRO. Blue and red spheres represent Weyl nodes with opposite chiralities ($\chi = +1$ and -1). The colored circles highlight one of two pairs of Weyl nodes closest to the Fermi level, both located at -13 meV. b, c) Three-dimensional energy dispersions of the (b) Weyl node pair and (c) nodal line, projected onto the $k_a$-$k_{\parallel,N}$ and $k_a$-$k_c$ planes, respectively. $k_{\parallel,N}$ is parallel to the direction connecting the Weyl node pair in (a). The black loop in (c) denotes the nodal line, which intersects the Fermi surface (gray) at eight distinct points. $k_{\parallel,L}$ denotes a path parallel to two of the eight crossing points. d, e) Band structures (top) of corresponding Berry curvature (bottom) along paths parallel to the (d) Weyl nodes and (e) nodal line. The circles indicate the positions of the topological nodes. f) Calculated AHC as a function of Fermi energy varying from -0.5 eV to 0.5 eV.

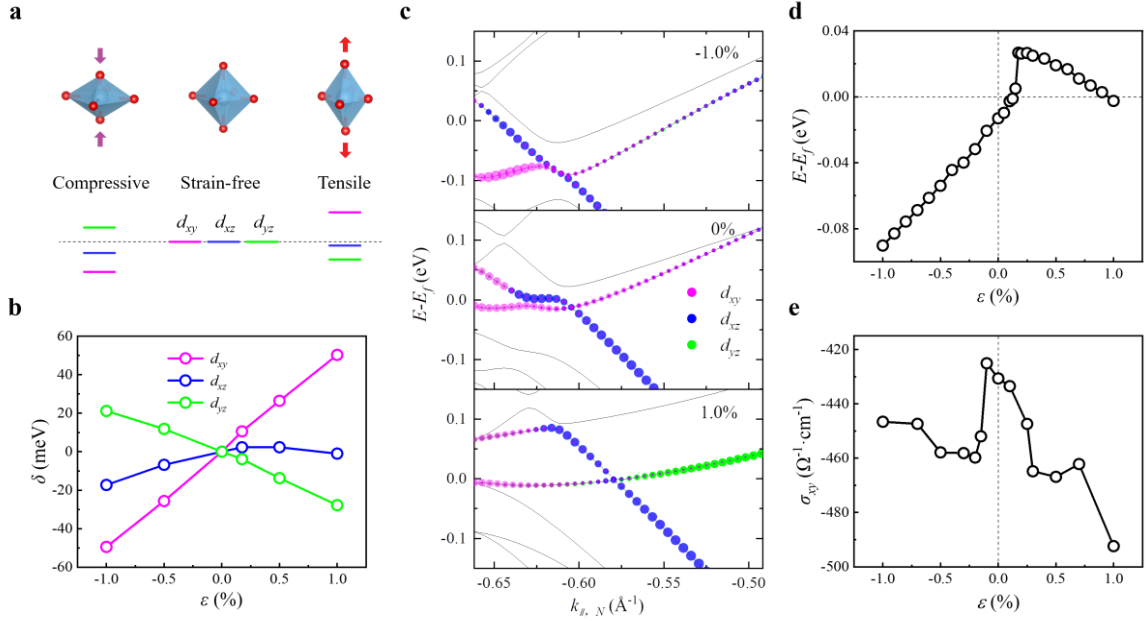

**Figure 4.** Evolution of Weyl nodes and AHC in SRO under strain. a) Schematic representation of oxygen octahedron deformation and $t_{2g}$ orbital shifts under tensile and compressive strains. Magenta, blue and green represent the $d_{xy}$, $d_{xz}$ and $d_{yz}$ orbitals, respectively. b) Relative energy splitting of the three $t_{2g}$ orbitals as a function of strain. c) $t_{2g}$ orbital-projected band structures of SRO under strains of -1.0%, 0% and +1.0%, with the color and size of the spheres representing the orbital composition and intensity, respectively. d) Evolution of the energy positions of Weyl nodes under different strains. e) Calculated non-monotonic variation of AHC as a function of strain.

WILEY-VCH

**Table of Contents**

We develop a flexural strain platform to isolate intrinsic strain effects on the topological electronic structure in SrRuO$_3$, and demonstrate the topological properties and associated anomalous Hall effect can be effectively modulated via strain engineering without introducing phase transitions or crystal defects, offering a new approach for designing flexible devices and optimizing performance based on topological oxides.

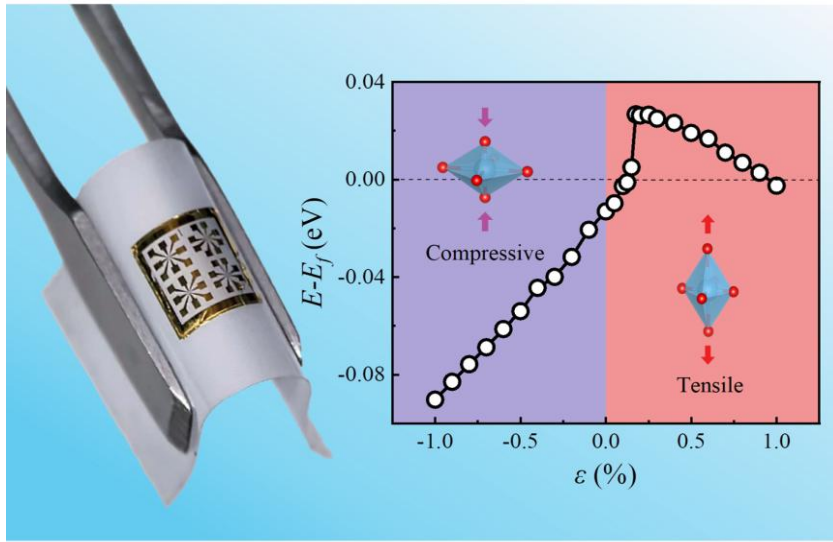